\documentclass[thmsa]{article}
\usepackage{sw20lart}

\input{tcilatex}

\begin{document}

\author{I. P\'{e}rez-Arjona, G.J. de Valc\'{a}rcel, and Eugenio Rold\'{a}n \\
Departament d'\`{O}ptica, Universitat de Val\`{e}ncia\\
Dr. Moliner 50, 46100--Burjassot, Spain}
\title{Two--Photon Absorption\\
Absorci\'{o}n de dos fotones}
\maketitle

\begin{abstract}
Two--photon absorption is theoretically analyzed within the semiclassical
formalism of radiation--matter interaction. We consider an ensemble of
inhomogeneously broadened three--level atoms subjected to the action of two
counterpropagating fields of the same frequency. By concentrating in the
limit of large detuning in one--photon transitions, we solve perturbatively
the Bloch equations in a non-usual way. In this way we derive an analytical
expression for the width of the two--photon resonance that makes evident
sub-Doppler two--photon spectroscopy. We also derive an analytical
expression for the Stark shift of the two--photon resonance.

PACS: 42.50.-p (Quantum Optics), 42.62.Fi (Laser Spectroscopy)
\end{abstract}

\begin{abstract}
Se analiza te\'{o}ricamente la absorci\'{o}n de dos fotones dentro del
formalismo semicl\'{a}sico de la interacci\'{o}n entre la radiaci\'{o}n y la
materia. Consideramos un conjunto, con ensanchamiento inhomog\'{e}neo, de 
\'{a}tomos de tres niveles sometido a la acci\'{o}n de dos campos
contrapropagantes de igual frequencia. Resolvemos perturbativamente las
ecuaciones de Bloch del sistema de una forma no usual concentr\'{a}ndonos en
el l\'{i}mite de alta desinton\'{i}a de las transiciones a un fot\'{o}n. De
esta forma obtenemos una expresi\'{o}n anal\'{i}tica para la anchura de la
resonancia a dos fotones en la que se pone de manifiesto la posibilidad de
espectroscop\'{i}a sub--Doppler a dos fotones. Tambi\'{e}n obtenemos una
expresi\'{o}n anal\'{i}tica para el desplazamiento Stark de la resonancia a
dos fotones.
\end{abstract}

\section{Introduction}

Two--photon absorption (TPA) is one of the most basic radiation--matter
interaction mechanisms. It consists in the excitation of an atom or molecule
from a lower quantum state $\left| 1\right\rangle $ to an excited state $%
\left| 2\right\rangle $ of the same parity as $\left| 1\right\rangle $ in a
single step. In this case the initial and final states cannot be connected
through an electric-dipole transition. Thus parity conservation implies that
two light quanta must be absorbed simultaneously. The theory of TPA was
first developed by Maria G\"{o}ppert--Mayer in 1931 in her Ph.D. Thesis \cite
{Göppert-Mayer31}.

As a multiphoton process, TPA is closely related to Raman scattering. In the
latter process, one photon is absorbed while the other is simultaneously
emitted, the energy difference being retained by the molecule. While
spontaneous Raman scattering was observed as early as 1928 \cite{Raman28},
TPA was not observed until 1961 \cite{Kaiser61} after the advent of the
laser (in fact TPA is one of the first nonlinear optical phenomena
demonstrated with the aid of laser radiation). The reason for that delay in
the observation of the two multiphoton processes lies in the fact that while
in spontaneous Raman scattering the scattered light intensity is
proportional to the intensity of the incoming radiation, in TPA the power
absorbed is proportional to the square of the intensity of the incoming
field and thus higher excitation energy is required for TPA.

TPA is a very important tool in laser spectroscopy as it makes possible the
transition between two states that cannot be connected by electric--dipole
interaction. Of course these transitions can also be investigated by making
use of resonant one--photon processes through an intermediate level, but in
this case the measured linewidth of the process is increased by the
linewidths of the two successive one--photon absorptions. TPA also allows
the coherent excitation of molecules to states whose energies fall in the
far ultraviolet, by making use of visible radiation, for which coherent
sources are easily available.

One of the most outstanding features of TPA is that it allows sub--Doppler
precision measurements \footnote{%
Raman scattering also allows the investigation of transitions in which the
initial and final states are of the same parity. With respect to Doppler
compensation, in Raman scattering it is only partial and the degree of
compensation depends on the energy difference between the initial and final
molecular states.}. This last fact was first analyzed by Vasilenko \textit{%
et al.} \cite{Vasilenko70} in 1970 and observed in 1974 \cite
{Cagnac74,Levenson74}. Doppler broadening comes from the fact that atoms
moving with different velocities ''see'' the field with different
frequencies because of the Doppler effect. This is a source of inhomogeneity
that increases the measured absorption linewidth. In one--photon transitions
this limitation cannot be easily overcome unless subtle phenomena such as
the Lamb-dip produced by spectral hole burning are exploited. In TPA,
however, there is a simple way of (almost) getting rid of Doppler
broadening. This occurs when the two photons inducing the transition come
from two counterpropagating beams of equal frequency. In this case all atoms
are in resonance with the two--photon process since the Doppler frequency
shifts of the two photons ''seen'' by the atom are opposite among them,
independent of the atom's velocity. Hence the sum of the energies of the two
photons, as ''seen'' by any atom, is twice the energy of a single photon in
the laboratory frame, and the inhomogeneity almost disappears.

In quantum optics textbooks, TPA is often introduced after field
quantization \cite{Loudon}. Nevertheless TPA does not need the existence of
photons to be understood and some textbooks analyze the phenomenon from a
semiclassical point of view \cite{Meystre} that is, by treating matter
quantum--mechanically and radiation classically (in this semiclassical
approach one must understand that the word photon refers to the amount of
energy absorbed by the atom, not to any quantum already existent in the free
electromagnetic field). There are several ways of studying TPA in this
semiclassical approach: derivation of nonlinear susceptibilities,
application of standard perturbation theory, even derivation of exact
analytical results. Nevertheless to our knowledge no standard textbook
derives the main characteristics of TPA (such as Doppler compensation and
the Stark shift of the resonance) within the semiclassical frame. In this
article we give a compact and clear presentation of TPA from a semiclassical
point of view, by solving perturbatively the equations of motion for the
density matrix elements.

\section{Semiclassical density matrix equations.}

Let us consider a classical monochromatic electromagnetic field of the form 
\begin{equation}
\mathbf{E}\left( z,t\right) =\mathbf{e}\left[ E_{1}\cos \left( \omega
t+kz\right) -E_{2}\cos \left( \omega t-kz\right) \right]
\end{equation}
where $\mathbf{e}$ is the unit polarization vector (linear polarization is
considered) and $E_{1}$ and $E_{2}$ are the constant real amplitudes of two
counterpropagating plane waves of angular frequency $\omega $ and wavenumber 
$k$, which travel along the $z$ axis. Note that this form of writing the
total field is completely general for the superposition of two
counterpropagating monochromatic linearly polarized waves of equal
polarization, since any dephasing between them can be removed by suitable
choice of time and space origins. This field represents a standing wave when 
$E_{1}=E_{2}$ and a traveling wave if either $E_{1}$ or $E_{2}$ is taken to
be zero.

This classical field interacts with a medium composed of three--level atoms
(Fig. 1): levels $\left| 1\right\rangle $ and $\left| 2\right\rangle $ of
the same parity, and contrary to that of the intermediate level $\left|
0\right\rangle $. This is the simplest level scheme that allows the
description of TPA in terms of the usual electric--dipole Hamiltonian. In
this way, the transition $\left| 1\right\rangle \longleftrightarrow \left|
2\right\rangle $ is produced via the virtual transitions $\left|
1\right\rangle \longleftrightarrow \left| 0\right\rangle $ and $\left|
0\right\rangle \longleftrightarrow \left| 2\right\rangle $ (state $\left|
0\right\rangle $ is kept far from resonance). The existence of an
intermediate level enhances the excitation probability between states $%
\left| 1\right\rangle $ and $\left| 2\right\rangle $ as will be shown.

The unperturbed hamiltonian $\widehat{H}_{0}$ of the three--level atoms is
given by (see level diagram in Fig.1) 
\begin{equation}
\widehat{H}_{0}=\hbar \left( \omega _{20}\left| 2\right\rangle \left\langle
2\right| -\omega _{01}\left| 1\right\rangle \left\langle 1\right| \right) ,
\end{equation}
and the origin of energies has been taken at the intermediate state $\left|
0\right\rangle $. Since levels $\left| 2\right\rangle $ and $\left|
1\right\rangle $ have the same parity, and contrary to that of state $\left|
0\right\rangle $, the allowed electric--dipole transitions are $\left|
1\right\rangle \longleftrightarrow \left| 0\right\rangle $ and $\left|
0\right\rangle \longleftrightarrow \left| 2\right\rangle $. Thus the
dipole--moment operator is written as 
\begin{equation}
\widehat{\mathbf{\mu }}=\mathbf{\mu }_{20}\left| 2\right\rangle \left\langle
0\right| +\mathbf{\mu }_{02}\left| 0\right\rangle \left\langle 2\right| +%
\mathbf{\mu }_{10}\left| 1\right\rangle \left\langle 0\right| +\mathbf{\mu }%
_{01}\left| 0\right\rangle \left\langle 1\right| ,
\end{equation}
where $\mathbf{\mu }_{ij}=\left\langle i\right| \widehat{\mathbf{\mu }}%
\left| j\right\rangle $, that can be taken to be real without loss of
generality through proper choice of the basis states phases $\left( \mathbf{%
\mu }_{ij}=\mathbf{\mu }_{ji}\right) $. The interaction hamiltonian of an
atom located at $z$ reads $\widehat{H}_{1}\left( z,t\right) =-\widehat{%
\mathbf{\mu }}\cdot \mathbf{E}\left( z,t\right) $ and the total hamiltonian
that governs the coherent evolution of the atoms is then $\widehat{H}_{S}=%
\widehat{H}_{0}+\widehat{H}_{1}$, where the subscript $S$ is used to denote
the Schr\"{o}dinger picture implicitly adopted. Before solving the Schr\"{o}%
dinger equation, it is convenient to remove fast oscillations at optical
frequencies appearing in the hamiltonian. This is accomplished by
transforming from the Schr\"{o}dinger picture to the field--interaction
picture. The appropriate unitary operator for making such transformation is 
\begin{equation}
\widehat{U}\left( t\right) =e^{i\omega t}\left| 2\right\rangle \left\langle
2\right| +\left| 0\right\rangle \left\langle 0\right| +e^{-i\omega t}\left|
1\right\rangle \left\langle 1\right| .
\end{equation}
Note that this operator is similar to that defining the Dirac picture but,
instead of removing the fast free atomic evolution (which would be
accomplished with the operator $\widehat{U}_{D}\left( t\right) =e^{i\omega
_{20}t}\left| 2\right\rangle \left\langle 2\right| +\left| 0\right\rangle
\left\langle 0\right| +e^{-i\omega _{01}t}\left| 1\right\rangle \left\langle
1\right| $), we remove the fast dynamics originating from the optical
frequency of the field. In the new picture, the state vector $\left| \psi
\right\rangle $ of the system $\left( \left| \psi \right\rangle =\widehat{U}%
\,\,\left| \psi _{S}\right\rangle \right) $ obeys the following Schr\"{o}%
dinger equation 
\[
i\hbar \frac{d}{dt}\left| \psi \right\rangle =\widehat{H}\,\,\left| \psi
\right\rangle \text{,}
\]
where the hamiltonian $\widehat{H}$ in the new picture\footnote{%
Under a unitary transformation, any operator $\widehat{O}_{S}$ (in the Schr%
\"{o}dinger picture) transforms according to the rule $\widehat{O}=\widehat{U%
}\widehat{O}_{S}\widehat{U}^{-1}$. Notice that $\widehat{H}$ is not
transformed in the same way.} is calculated \cite{Galindo} through 
\begin{equation}
\widehat{H}=\widehat{U}\,\widehat{H}_{S}\widehat{U}^{-1}+i\hbar \frac{%
\partial \widehat{U}}{\partial t}\widehat{U}^{-1}.
\end{equation}
After performing the rotating wave approximation \cite{Loudon,Meystre} (that
consists in removing fast oscillating terms) the hamiltonian reads 
\begin{equation}
\widehat{H}\left( z\right) =\hbar \left( -\delta _{2}\left| 2\right\rangle
\left\langle 2\right| +\delta _{1}\left| 1\right\rangle \left\langle
1\right| -\mu E\left| 2\right\rangle \left\langle 0\right| -E\left|
0\right\rangle \left\langle 1\right| +h.c.\right) ,  \label{hamiltonian}
\end{equation}
where 
\begin{eqnarray}
E &=&\phi _{1}e^{ikz}-\phi _{2}e^{-ikz}, \\
\phi _{1\left( 2\right) } &=&\frac{\mathbf{\mu }_{10}\cdot \mathbf{e}}{%
2\hbar }E_{1\left( 2\right) },\,\,\,\,\,\,\,\,\mu =\frac{\mathbf{\mu }%
_{20}\cdot \mathbf{e}}{\mathbf{\mu }_{10}\cdot \mathbf{e}},  \label{mu} \\
\delta _{1\left( 2\right) } &=&\omega -\omega _{01\left( 20\right) }.
\end{eqnarray}
and $h.c.$ stands for hermitian--conjugate. Note that the new picture, in
combination with the rotating wave approximation, yields a hamiltonian
independent of time. $E$ ($\mu E$) is half the complex Rabi frequency of the
field associated with the lower (upper) transition of an atom located at
position $z$.

Now we determine the evolution equation of the density matrix (more
properly: the population matrix, see below). We choose to work with the
density matrix instead of the state vector since in this way relaxation and
pumping processes can be (phenomenologically) incorporated into the model in
a simple way. As we are considering not a single molecule but a large number
of molecules which are moving at different velocities, an ensemble average
must be performed. The ensemble averaged density matrix is usually called
population matrix \cite{Sargent}. This ensemble must be defined for each
velocity and, since the interaction depends on space, the population matrix
must also be defined as a function of the position $z$: 
\begin{equation}
\widehat{\rho }\left( v;z,t\right) =\mathcal{N}\left( v\right) ^{-1}\tsum_{a}%
\widehat{\rho }_{a}\left( v;z,t\right) .
\end{equation}
Here $\widehat{\rho }$ is the population matrix, $\widehat{\rho }_{a}$ is
the density matrix for an atom labeled by $a$, and $a$ runs along all
molecules with velocity $v$ that, at time $t$, are within $z$ and $z+dz$ . $%
\mathcal{N}\left( v\right) $ is the number of such molecules, which is
assumed to be independent of $z$ and $t$ (homogeneity and stationarity of
the velocity distribution is assumed). The equation of evolution of the
population matrix is formally like the Schr\"{o}dinger--von Neumann equation
governing the evolution of the density matrix of a single atom, plus an
additional term \cite{Sargent}: 
\begin{equation}
\left( \partial _{t}+v\partial _{z}\right) \rho _{ij}=\left( i\hbar \right)
^{-1}\left[ \widehat{H},\widehat{\rho }\right] _{ij}+\left( \hat{\Gamma}%
\widehat{\rho }\right) _{ij}\text{,}  \label{vonneumann}
\end{equation}
$\left( i,j=0,1,2\right) $. $\hat{\Gamma}\widehat{\rho }$ describes
irreversible processes (relaxations and pumping) and $\hat{\Gamma}$ is a
generalized Liouvillian. In this article we shall consider the simple
expression 
\begin{equation}
\left( \hat{\Gamma}\widehat{\rho }\right) _{ij}=-\gamma \rho _{ij}+\gamma
\,\delta _{i,1}\delta _{j,1},  \label{liouville}
\end{equation}
with $\delta $ the Kronecker delta. The first contribution describes
relaxations in a situation in which all density matrix elements decay with
the same constant $\gamma $ (absence of dephasing collisions \cite{Sargent}%
). The second contribution (pump) guarantees that the ground state $\left|
1\right\rangle $ is asymptotically filled in the absence of interaction.
With this choice for $\hat{\Gamma}\widehat{\rho }$, $Tr\left( \hat{\rho}%
\right) =1$ always. We adopt this simple limit because the expressions are
much clearer and the details of the relaxation processes do not modify the
essential physics of TPA.

By substituting Eqs.(\ref{hamiltonian}) and (\ref{liouville}) into Eq.(\ref
{vonneumann}), the final equations of evolution of the population matrix
elements run 
\begin{eqnarray}
\left( \partial _{t}+v\partial _{z}\right) \rho _{22} &=&-\gamma \rho
_{22}+i\mu \left( E\rho _{02}-E^{\ast }\rho _{20}\right) ,  \label{mat1} \\
\left( \partial _{t}+v\partial _{z}\right) \rho _{00} &=&-\gamma \rho
_{00}+i\left( E\rho _{10}-E^{\ast }\rho _{01}\right) -i\mu \left( E\rho
_{02}-E^{\ast }\rho _{20}\right) ,  \label{mat2} \\
\left( \partial _{t}+v\partial _{z}\right) \rho _{11} &=&\gamma \left(
1-\rho _{11}\right) +i\left( E^{\ast }\rho _{01}-E\rho _{10}\right) ,
\label{mat3} \\
\left( \partial _{t}+v\partial _{z}\right) \rho _{21} &=&-\left( \gamma
-i\delta \right) \rho _{21}+iE\left( \mu \rho _{01}-\rho _{20}\right) ,
\label{mat4} \\
\left( \partial _{t}+v\partial _{z}\right) \rho _{20} &=&-\left( \gamma -i%
\frac{\delta -\Delta }{2}\right) \rho _{20}+i\mu E\left( \rho _{00}-\rho
_{22}\right) -iE^{\ast }\rho _{21},  \label{mat5} \\
\left( \partial _{t}+v\partial _{z}\right) \rho _{01} &=&-\left( \gamma -i%
\frac{\delta +\Delta }{2}\right) \rho _{01}+iE\left( \rho _{11}-\rho
_{00}\right) +i\mu E^{\ast }\rho _{21}.  \label{mat6}
\end{eqnarray}
where 
\begin{eqnarray}
\delta  &=&\delta _{1}+\delta _{2}=2\omega -\omega _{21}, \\
\Delta  &=&\delta _{1}-\delta _{2}=\omega _{20}-\omega _{01},
\end{eqnarray}
have the meaning of two--photon detuning and intermediate level detuning,
respectively (see Fig.1). The above equations should be complemented with
the evolution equation of the electromagnetic field. Nevertheless we shall
consider $E$ as a parameter. This corresponds to a physical situation in
which the gas of molecules is confined within a small region of the space
which is large compared with the radiation wavelength but small enough for
neglecting field depletion (thin film approximation).

Note that $\Delta $ is a structural parameter of the atoms, and we shall
consider only the case in which $\Delta $ is a very large quantity as
compared with the rest of frequencies $\left( \gamma ,\delta ,E\right) $
appearing in the problem. This limit guarantees that one--photon processes (%
\textit{i.e.} the electric--dipole transitions $\left| 1\right\rangle
\longleftrightarrow \left| 0\right\rangle $ and $\left| 0\right\rangle
\longleftrightarrow \left| 2\right\rangle $) are severely punished since the
one-photon detunings $\delta _{1}\left( \approx \Delta \right) $ and $\delta
_{2}\left( \approx -\Delta \right) $ are much larger than the widths of the
one-photon resonances. For example, consider the states $\left|
2\right\rangle =8S_{1/2}$, $\left| 0\right\rangle =7P$ and $\left|
1\right\rangle =6S_{1/2}$ of Cs. In this case \cite{Menshulach} $\omega
_{01}=4.098\cdot 10^{18}s^{-1}$ and $\omega _{20}=0.489\cdot 10^{18}s^{-1}$
and thus $\Delta =-3.609\cdot 10^{18}s^{-1}$. Cs is a gas and the
one--photon transitions width can be estimated to be given by their Doppler
width which, at room temperature are (see Section 4) $2.6\cdot 10^{8}s^{-1}$
and $2.21\cdot 10^{9}s^{-1}$ for the upper and lower transitions,
respectively: in this case there are nine orders of magnitude between $%
\Delta $ and the width of the one--photon resonances.

Eqs.(\ref{mat1})-(\ref{mat6}) do not admit a simple analytical solution in
the general case but can be solved perturbatively in the case of very large $%
\Delta $. This is done in the next section.

\section{Perturbative solution of the steady state}

In this section we solve perturbatively the equations of evolution of the
density matrix in steady state $\left( \partial _{t}\rightarrow 0\right) $.
Note that this is the state asymptotically reached by the system due to the
presence of relaxations. Here we present the main results and leave the
details to Appendix A. As commented, we shall consider the limit $\Delta \gg
\gamma ,\delta ,E$. We also consider that the inhomogeneous width $\gamma
_{v}$ (see Section 4) is small as compared with $\Delta $, that is, we
assume that $\Delta \gg kv$ with $k$ the field wavenumber. This can be made
formally explicit by writing $\Delta =\varepsilon ^{-1}\Delta _{1}$ with $%
\Delta _{1}$ a quantity of the same order of magnitude as the rest of the
frequencies present in the problem and $0<\varepsilon \ll 1$ (smallness
parameter). We also make a series expansion of the density matrix elements
of the form 
\begin{equation}
\rho _{ij}\left( z\right) =\sum_{n=0}^{\infty }\varepsilon ^{n}\rho
_{ij}^{\left( n\right) }\left( z\right) .
\end{equation}
Substituting this expansion in the population matrix equations and
identifying terms of equal powers in $\varepsilon $, one gets 
\begin{eqnarray}
0 &=&\left( v\partial _{z}+\gamma \right) \rho _{22}^{\left( n\right)
}+-i\,\mu \left( E\rho _{02}^{\left( n\right) }-E^{\ast }\rho _{20}^{\left(
n\right) }\right) , \\
0 &=&\left( v\partial _{z}+\gamma \right) \rho _{00}^{\left( n\right)
}-i\left( E\rho _{10}^{\left( n\right) }-E^{\ast }\rho _{01}^{\left(
n\right) }\right)  \\
&&+i\mu \left( E\rho _{02}^{\left( n\right) }-E^{\ast }\rho _{20}^{\left(
n\right) }\right) ,  \nonumber \\
0 &=&-\gamma +\left( v\partial _{z}+\gamma \right) \rho _{11}^{\left(
n\right) }+i\left( E\rho _{02}^{\left( n\right) }-E^{\ast }\rho
_{20}^{\left( n\right) }\right) , \\
0 &=&\left( v\partial _{z}+\gamma -i\delta \right) \rho _{21}^{(n)}-iE\left(
\mu \rho _{01}^{(n)}-\rho _{20}^{(n)}\right) , \\
-i\tfrac{1}{2}\Delta _{1}\rho _{20}^{\left( n+1\right) } &=&\left( v\partial
_{z}+\gamma -i\tfrac{1}{2}\delta \right) \rho _{20}^{\left( n\right) }-i\mu
E\left( \rho _{00}^{\left( n\right) }-\rho _{22}^{\left( n\right) }\right)
+iE^{\ast }\rho _{21}^{\left( n\right) }, \\
i\tfrac{1}{2}\Delta _{1}\rho _{01}^{\left( n+1\right) } &=&\left( v\partial
_{z}+\gamma -i\tfrac{1}{2}\delta \right) \rho _{01}^{\left( n\right)
}-iE\left( \rho _{11}^{\left( n\right) }-\rho _{00}^{\left( n\right)
}\right) -i\mu E^{\ast }\rho _{21}^{\left( n\right) },
\end{eqnarray}
where $n$ runs from $-1$ to $\infty $. Note that these equations refer to an
ensemble of atoms moving with velocity $v$ located at $z$. These equations
can be solved at each order $n$ of $\varepsilon $. We can integrate the
first four equations to obtain the populations $\rho _{ii}^{\left( n\right) }
$ and the two--photon coherence $\rho _{21}^{\left( n\right) }$ if we know
the value of the one photon coherences at this order ($\rho _{01}^{\left(
n\right) }$ and $\rho _{20}^{\left( n\right) }$). These quantities are
obtained from the two last equations. Note that the form of these last two
equations (which relate two consecutive orders) allow the values of the
one-photon coherences at a given order $n+1$ to be algebraically determined
in terms of the previous order $n$. In particular, for $n=-1$ we obtain $%
\rho _{01}^{\left( 0\right) }=$ $\rho _{20}^{\left( 0\right) }=0$, since $%
\rho _{ij}^{\left( -1\right) }=0$. These values allow to solve, from the
first four equations, for the rest of matrix elements at order $n=0$. Next, $%
\rho _{01}^{\left( 1\right) }$ and $\rho _{20}^{\left( 1\right) }$ are
determined from the last two equations and so on. There is just a single
point that deserves some explanation and concerns the integration in $z$ of
the first four equations. Notice that although we do not know any boundary
conditions (in terms of $z$) for the variables, we can make use of the
knowledge that, when the field is off $\left( E=0\right) $, all variables
must vanish at any order but $\rho _{11}^{(0)}$, which must be equal to
unity since $Tr\left( \hat{\rho}\right) =1$. In Appendix A the equations are
solved systematically. In the following we make use of the result of the
integration.

\section{Velocity and space averages}

We must concentrate on the calculation of a quantity directly related with
measurement. We shall consider the fluorescence intensity from the system,
which is directly proportional to the amount of population excited to the
upper level. The fluorescence signal collected by a detector will come from
all atoms (all velocities) existing within a finite region (of length $L$)
of the system. Thus it is necessary to perform both spatial and velocity
averages. The spatial average reads 
\begin{equation}
\left\langle \rho _{22}^{\left( n\right) }\left( v\right) \right\rangle _{z}=%
\frac{1}{L}\int_{0}^{L}dz\,\rho _{22}^{\left( n\right) }\left( v,z\right) ,
\label{espatialaverage}
\end{equation}
where we shall take, as already commented, $L\gg \lambda $ (with $\lambda $
the light wavelength) since typically the detector will collect the
fluorescence from a ''macroscopic'' region of the system. It is evident that
only the spatial dc component of $\rho _{22}^{\left( n\right) }\left(
v,z\right) $ will contribute to the spatial average (\ref{espatialaverage})
since $L\gg \lambda $, as stated. Consequently it will suffice to calculate
only those terms.

With respect to the velocity average a few words are in order. In a gas,
inhomogeneous broadening is due to the Doppler effect which is different for
each atomic velocity. The atomic velocities of a gas obey the
Maxwell--Boltzmann distribution 
\begin{equation}
\mathcal{G}\left( v\right) =\frac{1}{u\sqrt{\pi }}\exp \left[ -\left( \frac{v%
}{u}\right) ^{2}\right] =\frac{2k}{\gamma _{v}}\sqrt{\frac{\ln 2}{\pi }}\exp %
\left[ -\left( \frac{2kv}{\gamma _{v}/\sqrt{\ln 2}}\right) ^{2}\right] ,
\label{gauss}
\end{equation}
with $u$ the most probable velocity given by $u=\sqrt{2k_{B}T/m}$ ($k_{B}$
is Boltzmann's constant, $T$ is the absolute temperature, and $m$ is the
molecular mass). $\gamma _{v}=2\sqrt{\ln 2}ku$ is the inhomogeneous HWHM
(half--width at half maximum) in terms of the frequency $\Omega =2kv$ (the
factor $2$ is added for later convenience, since in TPA it is not the
radiation frequency -or its wavenumber- that is the important parameter but
twice its value). The problem with the Gaussian distribution is that some
integrals appearing in the final expressions cannot be evaluated
analytically. In order to obtain analytic expressions as simple as possible,
we shall consider a Lorentzian distribution for the atomic velocities 
\begin{equation}
\mathcal{L}(v)=\frac{1}{\pi }\frac{u}{u^{2}+v^{2}}=\frac{2k}{\pi }\frac{%
\gamma _{v}}{\gamma _{v}^{2}+\left( 2kv\right) ^{2}},  \label{lorentzian}
\end{equation}
where $\gamma _{v}=2ku$ is the inhomogeneous HWHM (half--width at half
maximum) in terms of the frequency $\Omega =2kv$. The results obtained with
this distribution will differ quantitatively but not qualitatively from the
Gaussian distribution, as will be shown.

The averaged population of the excited level is then calculated through 
\begin{equation}
\left\langle \rho _{22}^{\left( n\right) }\right\rangle =\int_{-\infty
}^{+\infty }dv\left\langle \rho _{ij}^{\left( n\right) }\left( v\right)
\right\rangle _{z}\mathcal{L}\left( v\right) .  \label{velocitysum}
\end{equation}
Clearly the averaging order is unimportant. We could first perform the
velocity average and then the spatial average, obtaining the same result.
From the computational viewpoint however it is more convenient to perform
first the spatial average since in this way the ac-components (in terms of $z
$) of $\rho _{22}^{\left( n\right) }$ are removed from the calculations from
the beginning.

From Eqs.(\ref{rho222}) and (\ref{rho223}) of Appendix B, the fully averaged
population of the upper level reads, up to order $\varepsilon ^{3}$, 
\[
\left\langle \rho _{22}\right\rangle =\varepsilon ^{2}\left\langle \rho
_{22}^{\left( 2\right) }\right\rangle +\varepsilon ^{3}\left\langle \rho
_{22}^{\left( 3\right) }\right\rangle \equiv N_{2}+N_{3} 
\]
where

\begin{equation}
N_{2}=8\mu ^{2}\left( \frac{\phi ^{2}}{\gamma \Delta }\right) ^{2}\left[ 
\frac{\left( 1+\tilde{\gamma}_{v}\right) (1+A^{4})}{\left( 1+\tilde{\gamma}%
_{v}\right) ^{2}+\tilde{\delta}^{2}}+\frac{4A^{2}}{1+\tilde{\delta}^{2}}%
\right] ,  \label{n2}
\end{equation}
and

\begin{eqnarray}
N_{3} &=&16\mu ^{2}\left( \mu ^{2}-1\right) \left( 1+A^{2}\right) \tilde{%
\delta}\left( \frac{\phi ^{2}}{\gamma \Delta }\right) ^{3}\left( \mathcal{B}%
_{1}+\mathcal{B}_{2}\right)   \label{n3} \\
\mathcal{B}_{1} &=&A^{2}\left[ \frac{2}{\left( 1+\tilde{\delta}^{2}\right)
^{2}}+\frac{1}{\tilde{\gamma}_{v}}\left( \frac{1}{1+\tilde{\delta}^{2}}-%
\frac{1}{\left( 1+\tilde{\gamma}_{v}\right) ^{2}+\tilde{\delta}^{2}}\right) %
\right]   \nonumber \\
\mathcal{B}_{2} &=&2\left( 1+A^{4}\right) \frac{\left( 1+\tilde{\gamma}%
_{v}\right) }{\left[ \left( 1+\tilde{\gamma}_{v}\right) ^{2}+\tilde{\delta}%
^{2}\right] ^{2}}.  \nonumber
\end{eqnarray}
In writing Eqs.(\ref{n2}) and (\ref{n3}) we have introduced the notation
\begin{equation}
\phi _{1}\equiv \phi ,\,\ \ \ \,\,\phi _{2}\equiv A\phi ,
\end{equation}
and the normalized frequencies
\begin{equation}
\tilde{\gamma}_{v}\equiv \frac{\gamma _{v}}{\gamma },\,\ \ \tilde{\delta}%
\equiv \frac{\delta }{\gamma }.
\end{equation}
Note that $\varepsilon ^{n}$ combines with $\Delta _{1}^{-n}$ in both orders 
$n=2$ and $n=3$ to yield $\Delta ^{-n}$, leading to a final expression
independent of $\varepsilon $. Next we analyze these expressions.

\section{Analysis of the results}

\subsection{Strength and width of the resonance}

In order to analyze the strength and width of the resonance it is suffice to
consider the dominant term $N_{2}$. General results are: (i) TPA is
proportional to the squared field intensity $\left( \phi ^{2}\right) ^{2}$,
(ii) The existence of an intermediate level with a finite detuning $\Delta $
enhances the probability of the process (the smaller $\Delta $ the larger
amount of excited population), and (iii) The maximum transfer of population
is produced at $\tilde{\delta}=0$ (this result will be corrected at the next
order; see next subsection). Let us consider some special cases.

In the case of homogeneous broadening $\left( \tilde{\gamma}_{v}=0\right) $,

\begin{equation}
N_{2}^{\hom }=8\mu ^{2}\left( \frac{\phi ^{2}}{\gamma \Delta }\right) ^{2}%
\frac{A^{4}+4A^{2}+1}{1+\tilde{\delta}^{2}}.  \label{pobhom}
\end{equation}
Note that $N_{2}^{\hom }$ is proportional to $(A^{4}+4A^{2}+1)$, which in
its turn is proportional to the mean value of the squared field intensity -a
signature of two--photon absorption. This factor is six times larger for
standing waves than for traveling waves. Note that this (important)
numerical factor is the single difference between standing and traveling
wave configurations in this homogeneous broadening limit. We conclude that,
from an experimental point of view, it is most convenient to illuminate the
cell with a traveling wave and make it reflect on a mirror located after the
cell in order to produce a standing wave. This represents no extra energetic
cost and the fluorescence signal collected in this way is 6 times larger
than without the mirror.

With a non-null inhomogeneous broadening two limits of interest are: a)
excitation with a traveling wave ($A=0$) 
\begin{equation}
N_{2}^{TW}=8\mu ^{2}\left( \frac{\phi ^{2}}{\gamma \Delta }\right) ^{2}\frac{%
\left( 1+\tilde{\gamma}_{v}\right) }{\left( 1+\tilde{\gamma}_{v}\right) ^{2}+%
\tilde{\delta}^{2}}{,}  \label{TW}
\end{equation}
and b) excitation with a standing wave ($A=1$) 
\begin{equation}
N_{2}^{SW}=8\mu ^{2}\left( \frac{\phi ^{2}}{\gamma \Delta }\right) ^{2}\left[
\frac{4}{\left( 1+\tilde{\delta}^{2}\right) }+\frac{2\left( 1+\tilde{\gamma}%
_{v}\right) }{\left( 1+\tilde{\gamma}_{v}\right) ^{2}+\tilde{\delta}^{2}}%
\right] {.}
\end{equation}
Note that the effect of the inhomogeneous broadening is dramatically
different for traveling wave or for standing wave cases: if $\tilde{\gamma}%
_{v}\gg 1$ (\textit{i.e.} $\gamma _{v}\gg \gamma $, inhomogeneous limit) $%
N_{2}^{TW}\rightarrow 0$, whereas in the same limit $N_{2}^{SW}\rightarrow 
\frac{2}{3}N_{2}^{\hom ,SW}$, where $N_{2}^{\hom ,SW}$ is obtained from $%
N_{2}^{\hom }$ by putting $A=1$.

In order to make clearer comparisons among different cases we next analyze
next the maximum of $N_{2}$ (that occurs at $\tilde{\delta}=0$ as stated)
and its width in terms of $\tilde{\delta}$.

The maximum of $N_{2}$ reads 
\begin{equation}
N_{2}^{\max }=8\mu ^{2}\left( \frac{\phi ^{2}}{\gamma \Delta }\right) ^{2}%
\left[ \frac{(A^{4}+4A^{2}+1)+4\tilde{\gamma}_{v}A^{2}}{\left( 1+\tilde{%
\gamma}_{v}\right) }\right] .
\end{equation}

In Fig. 2 $N_{2}^{\max }$ (normalized to its maximum value, for $A=1$ and $%
\tilde{\gamma}_{v}=0$) is plotted as a function of the
inhomogeneous-to-homogeneous widths ratio $\tilde{\gamma}_{v}$ for different
values of $A$. Clearly, for $A=1$ (standing wave) TPA is almost insensitive
to the amount of inhomogeneous broadening, whereas for $A=0$ (travelling
wave) the decrease in TPA is dramatic for ratios as moderate as $\tilde{%
\gamma}_{v}=1$ or larger.

As a function of the normalized detuning $\tilde{\delta}$, $N_{2}$ has a
bell shape whose FWHM $\Gamma $ is easily calculated from Eq.(\ref{n2}), and
reads 
\begin{eqnarray}
\Gamma ^{2} &=&4\left[ \sqrt{w+\left( w-1\right) ^{2}f^{2}}+\left(
w-1\right) f\right] ,  \label{FWHM} \\
w &=&\left( 1+\tilde{\gamma}_{v}\right) ^{2}, \\
f &=&\frac{1}{2}\frac{1+A^{4}-4\left( 1+\tilde{\gamma}_{v}\right) A^{2}}{%
1+A^{4}+4\left( 1+\tilde{\gamma}_{v}\right) A^{2}}.
\end{eqnarray}
For a homogeneously broadened line $\left( \tilde{\gamma}_{v}=0:w=1\right) $
the width reads $\Gamma _{\hom }=2$ (\textit{i.e.} in terms of the detuning $%
\delta $ the width reads $2\gamma $). In the special case of a traveling
wave $\left( A=0:f=1/2\right) $ the width reads $\Gamma ^{TW}=2\left( 1+%
\tilde{\gamma}_{v}\right) $, \textit{i.e.}, the width is the sum of the
homogeneous and inhomogeneous widths. For a standing wave $\left( A=1\right) 
$ no simple expression is obtained; nevertheless in the special case of
large inhomogeneous broadening $\left( \tilde{\gamma}_{v}\gg 1:f\rightarrow
-1/2,w\rightarrow \infty \right) $ the width reads $\Gamma ^{SW}\left( 
\tilde{\gamma}_{v}\rightarrow \infty \right) =2\left( 1+1/2\tilde{\gamma}%
_{v}\right) $ which tends to the homogeneous width $2$ for sufficiently
large inhomogeneous broadening. This last result is a fundamental property
of TPA: sub--Doppler spectroscopy can be performed in TPA experiments by
using a standing wave \cite
{Göppert-Mayer31,Kaiser61,Vasilenko70,Cagnac74,Levenson74,Bloembergen76}.

In Fig.3 we plot $\Gamma /_{\hom }$, as given by Eq.(\ref{FWHM}), as a
function of the normalized inhomogeneous width $\tilde{\gamma}_{v}$ for $A=1$
and $A=0.5$. Clearly, for any value of $A$ different from zero, $\Gamma
/_{\hom }$ first grows until it reaches a maximum and finally decreases
tending to unity for large enough $\tilde{\gamma}_{v}$. Of course the
optimum situation corresponds to $A=1\footnote{$A\neq 1$ can be understood
as the sum of a traveling wave and a standing wave. Thus the result in that
case is the sum of the two contributions. As the $TW$ contribution is less
important the larger is $\gamma _{v}$ and the contribution of the $SW$ is
basically independent of $\gamma _{v}$ this explains the above result. The
main difference between $A=1$ and $A\neq 1$ lies in the strength of the
resonance as shown in Fig.2.}$. Thus for large enough $\tilde{\gamma}_{v}$
the inhomogeneous broadening does not contribute at all to the width of the
resonance. 

In Fig.4 we show the same representation for $A=1$ (full line) together with
the numerical integration assuming a Gaussian velocity distribution. It can
be seen that the dependence is qualitatively the same and that only
relatively small numerical deviations are appreciated between both cases.
This confirms that the exact form of the velocity distribution is not very
important, whenever it is bell shaped.

\subsection{Shift of the resonance}

As we have seen, at order $\varepsilon ^{2}$ the maximum of the resonance is
located at $\tilde{\delta}=0$. Nevertheless, two--photon processes induce a
shift of the resonance, the so called Stark shift. This shift is only
captured at third order of the perturbative expansion. Making use of Eqs.(%
\ref{n2}) and (\ref{n3}), we compute $\partial \left( N_{2}+N_{3}\right)
/\partial \tilde{\delta}=0$ and obtain

\begin{equation}
\tilde{\delta}_{Stark}=2\left( 1+A^{2}\right) \left( \mu ^{2}-1\right)
\left( \frac{\phi ^{2}}{\gamma \Delta }\right) \frac{(1+A^{4})+A^{2}\left( 1+%
\tilde{\gamma}_{v}\right) \left( 2+5\tilde{\gamma}_{v}/2+\tilde{\gamma}%
_{v}^{2}\right) }{\left( 1+A^{4}\right) +4A^{2}\left( 1+\tilde{\gamma}%
_{v}\right) ^{3}}  \label{stark}
\end{equation}
which is the Stark shift. Note that this shift is proportional to $\left(
\phi ^{2}/\gamma \Delta \right) $, and is thus of order $\varepsilon $. Note
also that whenever $\mu =1$ (\textit{i.e.} when both one-photon transitions
have equal electric dipole matrix elements, see Eq.(\ref{mu})) the shift
vanishes. We see that the sign of the shift depends both on the sign of the
intermediate level detuning $\Delta $ and on the asymmetry between both
one--photon transitions through the quantity $\left( \mu ^{2}-1\right) $.

Particular cases of interest are: a) excitation with a traveling wave ($A=0$)

\begin{equation}
\tilde{\delta}_{Stark}^{TW}=2\left( \mu ^{2}-1\right) \left( \frac{\phi ^{2}%
}{\gamma \Delta }\right) ,
\end{equation}
which is independent of the inhomogeneous broadening, and b) excitation with
a standing wave ($A=1$)

\begin{equation}
\tilde{\delta}_{Stark}^{SW}=\left( \mu ^{2}-1\right) \left( \frac{\phi ^{2}}{%
\gamma \Delta }\right) \left[ 1+\frac{5+3\tilde{\gamma}_{v}+\tilde{\gamma}%
_{v}^{2}}{1+2\left( 1+\tilde{\gamma}_{v}\right) ^{3}}\right] 
\end{equation}
which tends to $\tilde{\delta}_{Stark}^{TW}/2$ for $\tilde{\gamma}%
_{v}\rightarrow \infty $. In Fig.5 $\tilde{\delta}_{Stark}^{SW}/\tilde{\delta%
}_{Stark}^{TW}$ is represented as a function of the inhomogeneous width for
both Lorentzian broadening (Eq.(\ref{stark})) and Gaussian broadening. Again
it can be appreciated that the results are very similar for both types of
inhomogeneous broadening.

\section{Conclusion}

In this article we have analytically studied two--photon absorption (TPA) in
an inhomogeneously broadened medium pumped by two counterpropagating light
beams of equal frequency. By making use of perturbative techniques, we have
derived explicit analytical expressions for the strength and width of the
resonance as well as for the Stark shift in the case of Lorentzian
broadening. Comparison with Gaussian broadening (numerically computed) has
shown that the qualitative features of TPA are quite independent of the
specific type of inhomogeneous broadening.

\section{Appendix A}

At order $\varepsilon ^{-1}$ one trivially gets 
\begin{equation}
\rho _{01}^{\left( 0\right) }=\rho _{20}^{\left( 0\right) }=0.
\label{orden0a}
\end{equation}

At order $\varepsilon ^{0}$ the equations are 
\begin{eqnarray}
v\partial _{z}\rho _{22}^{(0)} &=&-\gamma \rho _{22}^{(0)}, \\
v\partial _{z}\rho _{00}^{(0)} &=&-\gamma \rho _{00}^{(0)}, \\
v\partial _{z}\rho _{11}^{(0)} &=&1-\gamma \rho _{11}^{(0)}, \\
v\partial _{z}\rho _{21}^{(0)} &=&-\left( \gamma -i\delta \right) \rho
_{21}^{(0)}, \\
v\partial _{z}\rho _{20}^{(0)} &=&-\frac{i}{2}\Delta \rho _{20}^{(1)}+iE\mu
\left( \rho _{00}^{(0)}-\rho _{22}^{(0)}\right) -iE^{\ast }\rho _{21}^{(0)},
\\
v\partial _{z}\rho _{01}^{(0)} &=&+\frac{i}{2}\Delta \rho
_{01}^{(1)}+iE\left( \rho _{11}^{(0)}-\rho _{00}^{(0)}\right) -iE^{\ast }\mu
\rho _{21}^{(0)},
\end{eqnarray}
whose solution is 
\begin{eqnarray}
\rho _{11}^{(0)}\left( v,z\right) &=&1,  \label{orden0b} \\
\rho _{22}^{(0)}\left( v,z\right) &=&\rho _{00}^{(0)}\left( v,z\right) =\rho
_{21}^{(0)}\left( v,z\right) =0,
\end{eqnarray}
and 
\begin{eqnarray}
\rho _{{01}}^{({1)}}\left( v,z\right) &=&-\frac{2\,}{\Delta }\left( \phi
_{1}e^{ikz}-\phi _{2}e^{-ikz}\right) ,  \label{orden1a} \\
\,\rho _{{20}}^{({1)}}\left( v,z\right) &=&0.
\end{eqnarray}
At order $\varepsilon ^{1}$ the equations are 
\begin{eqnarray}
v\partial _{z}\rho _{22}^{(1)} &=&-\gamma \rho _{22}^{(1)}+i\mu \left( E\rho
_{02}^{(1)}-E^{\ast }\rho _{20}^{(1)}\right) , \\
v\partial _{z}\rho _{00}^{(1)} &=&-\gamma \rho _{00}^{(1)}+i\left( E\rho
_{10}^{(1)}-E^{\ast }\rho _{01}^{(1)}\right) -i\mu \left( E\mu \rho
_{02}^{(1)}-E^{\ast }\rho _{20}^{(1)}\right) , \\
v\partial _{z}\rho _{11}^{(1)} &=&-\gamma \rho _{11}^{(1)}+i\left( E^{\ast
}\rho _{01}^{(1)}-E\rho _{10}^{(1)}\right) , \\
v\partial _{z}\rho _{21}^{(1)} &=&-\left( \gamma -i\delta \right) \rho
_{21}^{(1)}+iE\left( \mu \rho _{01}^{(1)}-\rho _{20}^{(1)}\right) , \\
v\partial _{z}\rho _{20}^{(1)} &=&-\left( \gamma -\frac{i}{2}\delta \right)
\rho _{20}^{(1)}-\frac{i}{2}\Delta _{1}\rho _{20}^{(2)}+iE\mu \left( \rho
_{00}^{(1)}-\rho _{22}^{(1)}\right) -iE^{\ast }\rho _{21}^{(1)}, \\
v\partial _{z}\rho _{01}^{(1)} &=&-\left( \gamma -\frac{i}{2}\delta \right)
\rho _{01}^{(1)}+\frac{i}{2}\Delta _{1}\rho _{01}^{(2)}+iE\left( \rho
_{11}^{(1)}-\rho _{00}^{(1)}\right) -iE^{\ast }\mu \rho _{21}^{(1)},
\end{eqnarray}
and integration along $z$ has to be carried out. By using Eqs.(\ref{orden1a}%
), it is straightforward to obtain that 
\begin{eqnarray}
\rho _{ii}^{\left( 1\right) }\left( v,z\right) &=&0,\,\,i=0,1,2
\label{orden1b1} \\
\rho _{21}^{\left( 1\right) }\left( v,z\right) &=&-\frac{2i\mu }{\Delta _{1}}%
\left[ \frac{\phi _{1}^{2}}{D_{+}}e^{2ikz}-\frac{2\phi _{1}\phi _{2}}{D_{0}}+%
\frac{\phi _{2}^{2}}{D_{-}}e^{-2\,i\,k\,z}\right] ,  \label{orden1b2}
\end{eqnarray}
and 
\begin{eqnarray}
\rho _{20}^{\left( 2\right) }\left( v,z\right) &=&\frac{4i\mu }{\Delta
_{1}^{2}}{\Large [}-\frac{\phi _{1}^{2}\phi _{2}}{D_{+}}e^{3ikz}+\left( 
\frac{\phi _{1}^{3}}{D_{+}}+\frac{2\phi _{1}\phi _{2}^{2}}{D_{0}}\right)
e^{ikz}-  \label{aux0} \\
&&-\left( \frac{\phi _{2}^{3}}{D_{-}}+\frac{2\phi _{1}^{2}\phi _{2}}{D_{0}}%
\right) e^{-ikz}+\frac{\phi _{1}\phi _{2}^{2}}{D_{-}}e^{-3ikz}{\Large ]}, 
\nonumber \\
\rho _{01}^{\left( 2\right) }\left( v,z\right) &=&\frac{4i\mu ^{2}}{\Delta
_{1}^{2}}{\Large \{}-\frac{\phi _{1}^{2}\phi _{2}}{D_{+}}e^{3\,i\,k\,z}+%
\left[ \frac{\left( \gamma +D_{+}\right) \phi _{1}}{2{\mu ^{2}}}+\left( 
\frac{\phi _{1}^{3}}{D_{+}}+\frac{2\phi _{1}{\phi }_{2}^{2}}{D_{0}}\right) %
\right] e^{ikz}  \nonumber  \label{orden2a2} \\
&&-\left[ \frac{\left( \gamma +D_{-}\right) \phi _{2}}{2{\mu ^{2}}}+\left( 
\frac{\phi _{2}^{3}}{D_{-}}+\frac{2{{\phi }_{1}^{2}\phi }_{2}}{D_{0}}\right) %
\right] e^{-ikz}+\frac{\phi _{1}\phi _{2}^{2}}{D_{-}}e^{-3i\,k\,z}{\Large \},%
}
\end{eqnarray}
with 
\begin{eqnarray}
D_{\pm } &=&\gamma -i\left( \delta \mp 2kv\right) , \\
D_{0} &=&\gamma -i\delta .
\end{eqnarray}

At order $\varepsilon ^{2}$ it is not necessary to compute all the terms
since we are only interested in $\rho _{22}^{\left( 2\right) }$ and $\rho
_{20}^{\left( 3\right) }$ (the latter is necessary for calculating $\rho
_{22}^{\left( 3\right) }$ at order $\varepsilon ^{3}$)\footnote{%
Notice that if one is interested only in the analysis of the strength and
width of the resonance (and not of the Stark shift), it is enough to
calculate the non-oscillating term in Eq.(\ref{aux2}) by direct substitution
of (\ref{aux0}) in Eq.(\ref{aux1}), quite a simple task. The rest of the
terms are necessary for obtaning of $\rho _{22}^{(3)}$ which becomes a
simple but tedious task.}. The necessary equations are 
\begin{eqnarray}
v\partial _{z}\rho _{22}^{\left( 2\right) } &=&-\gamma \rho _{22}^{\left(
2\right) }+i\,\mu \left( E\rho _{02}^{\left( 2\right) }-E^{\ast }\rho
_{20}^{\left( 2\right) }\right) ,  \label{aux1} \\
v\partial _{z}\rho _{00}^{\left( 2\right) } &=&-\,\gamma \rho _{00}^{\left(
2\right) }+i\left( E\rho _{10}^{\left( 2\right) }-E^{\ast }\rho
_{01}^{\left( 2\right) }\right) -i\mu \left( E\rho _{02}^{\left( 2\right)
}-E^{\ast }\rho _{20}^{\left( 2\right) }\right) , \\
v\partial _{z}\rho _{21}^{\left( 2\right) } &=&-\left( \gamma -i\delta
\right) \rho _{21}^{(2)}+iE\left( \mu \rho _{01}^{(2)}-\rho
_{20}^{(2)}\right) , \\
v\partial _{z}\rho _{20}^{\left( 2\right) } &=&-\left( \gamma -i\frac{\delta 
}{2}\right) \rho _{20}^{\left( 2\right) }-i\frac{\Delta _{1}}{2}\rho
_{20}^{\left( 3\right) }+i\mu E\left( \rho _{00}^{\left( 2\right) }-\rho
_{22}^{\left( 2\right) }\right) -iE^{\ast }\rho _{21}^{\left( 2\right) },
\end{eqnarray}
and the searched quantities are given by 
\begin{eqnarray}
\rho _{22}^{\left( 2\right) }\left( v,z\right)  &=&\frac{4\mu ^{2}}{\gamma
\Delta _{1}^{2}}\left[ \frac{\phi _{1}^{4}}{D_{+}}+\frac{\phi _{2}^{4}}{D_{-}%
}+\frac{4\phi _{1}^{2}\phi _{2}^{2}}{D_{0}}\right] +c.c.+  \label{aux2} \\
&&-\frac{16\mu ^{2}\phi _{1}\phi _{2}}{\Delta _{1}^{2}}\left( \frac{\gamma
+ikv}{\gamma -2ikv}\right) \left[ \frac{\phi _{1}^{2}}{D_{0}^{\ast }D_{+}}+%
\frac{\phi _{2}^{2}}{D_{0}D_{-}^{\ast }}\right] e^{i2kz}+c.c.+  \nonumber \\
&&+terms\,\,with\,\,e^{\pm i4kz}\text{ },  \nonumber \\
\rho _{00}^{\left( 2\right) }\left( v,z\right)  &=&\frac{8}{\Delta _{1}^{2}}%
\left[ \phi _{1}^{2}+\phi _{2}^{2}-\frac{\gamma +ikv}{\gamma +2ikv}\phi
_{1}\phi _{2}e^{i2kz}-c.c.\right] + \\
&&+terms\,\,with\,\,e^{inkz}\text{ }\left( \text{\thinspace }n\neq 0,\pm
2\right) ,  \nonumber \\
\rho _{21}^{\left( 2\right) }\left( v,z\right)  &=&\frac{4\mu \phi _{1}\phi
_{2}}{\Delta _{1}^{2}D_{0}}\left[ \gamma +D_{0}+\left( \mu ^{2}-1\right)
\left( 2\frac{\phi _{1}^{2}+\phi _{2}^{2}}{D_{0}}+\frac{\phi _{1}^{2}}{D_{+}}%
+\frac{\phi _{2}^{2}}{D_{-}}\right) \right] - \\
&&-\frac{2\mu \phi _{1}^{2}}{\Delta _{1}^{2}D_{0}}\left[ \gamma
+D_{+}+2\left( \mu ^{2}-1\right) \left( 2\frac{\phi _{2}^{2}}{D_{0}}+\frac{%
\phi _{1}^{2}+\phi _{2}^{2}}{D_{-}}\right) \right] e^{i2kz}-  \nonumber \\
&&-\frac{2\mu \phi _{2}^{2}}{\Delta _{1}^{2}D_{0}}\left[ \gamma
+D_{-}+2\left( \mu ^{2}-1\right) \left( 2\frac{\phi _{1}^{2}}{D_{0}}+\frac{%
\phi _{1}^{2}+\phi _{2}^{2}}{D_{+}}\right) \right] e^{-i2kz}+  \nonumber \\
&&+terms\,\,with\,\,e^{inkz}\text{ }\left( \text{\thinspace }n\neq 0,\pm
2\right)   \nonumber
\end{eqnarray}
from the three first equations and, from the last equation, 
\begin{equation}
\rho _{20}^{\left( 3\right) }\left( v,z\right) =\rho _{20}^{\left(
3,+\right) }e^{ikz}+\rho _{20}^{\left( 3,-\right)
}e^{-ikz}+terms\,\,with\,\,e^{inkz},\text{ }n\neq \pm 1
\end{equation}
where 
\begin{eqnarray}
\rho _{20}^{\left( 3,+\right) } &=&\frac{8\mu \phi _{1}}{\Delta _{1}^{3}}%
{\Huge \{}\left[ \frac{\left( \mu ^{2}-1\right) }{D_{+}^{2}}-\frac{4\mu ^{2}%
}{\left| D_{+}\right| ^{2}}\right] \phi _{1}^{4}+2\phi _{1}^{2}+  \nonumber
\\
&&+\left[ \frac{\left( \mu ^{2}-1\right) }{D_{0}}\left( \frac{2}{D_{0}}+%
\frac{3}{D_{+}}+\frac{D_{0}}{D_{+}^{2}}\right) -\frac{4\mu ^{2}}{D_{0}^{\ast
}}\left( \frac{4}{D_{0}}-\frac{\left( \gamma +ikv\right) }{D_{+}\left(
\gamma -2ikv\right) }\right) \right] \phi _{1}^{2}\phi _{2}^{2}+  \nonumber
\\
&&+\left[ 4-\frac{D_{+}}{D_{0}}+\frac{\gamma }{\gamma +2ikv}\right] \phi
_{2}^{2}+  \nonumber \\
&&+\left[ \frac{\left( \mu ^{2}-1\right) }{D_{0}}\left( \frac{1}{D_{-}}+%
\frac{2}{D_{0}}\right) -\frac{4\mu ^{2}}{D_{-}^{\ast }}\left( \frac{1}{D_{-}}%
+\frac{\left( \gamma +ikv\right) }{D_{0}\left( \gamma -2ikv\right) }\right) 
\right] \phi _{2}^{4}{\Huge \}},
\end{eqnarray}
and 
\begin{eqnarray}
\rho _{20}^{\left( 3,-\right) } &=&-\frac{8\mu \phi _{2}}{\Delta _{1}^{3}}%
{\Huge \{}\left[ \frac{\left( \mu ^{2}-1\right) }{D_{-}^{2}}-\frac{4\mu ^{2}%
}{\left| D_{-}\right| ^{2}}\right] \phi _{2}^{4}+2\phi _{2}^{2}+  \nonumber
\\
&&+\left[ \frac{\left( \mu ^{2}-1\right) }{D_{0}}\left( \frac{2}{D_{0}}+%
\frac{3}{D_{-}}+\frac{D_{0}}{D_{-}^{2}}\right) -\frac{4\mu ^{2}}{D_{0}^{\ast
}}\left( \frac{4}{D_{0}}+\frac{\left( \gamma -ikv\right) }{D_{-}\left(
\gamma +2ikv\right) }\right) \right] \phi _{2}^{2}\phi _{1}^{2}+  \nonumber
\\
&&+\left[ 4-\frac{D_{-}}{D_{0}}+\frac{\gamma }{\gamma -2ikv}\right] \phi
_{1}^{2}+  \nonumber \\
&&+\left[ \frac{\left( \mu ^{2}-1\right) }{D_{0}}\left( \frac{1}{D_{+}}+%
\frac{2}{D_{0}}\right) -\frac{4\mu ^{2}}{D_{+}^{\ast }}\left( \frac{1}{D_{+}}%
-\frac{\left( \gamma -ikv\right) }{D_{0}\left( \gamma +2ikv\right) }\right) 
\right] \phi _{1}^{4}{\Huge \}}.
\end{eqnarray}

Finally, at order $\varepsilon ^{3}$ we are only interested in obtaining the
value of the population of the excited level. Thus we only need

\begin{equation}
v\partial _{z}\rho _{22}^{(3)}=\gamma \rho _{22}^{(3)}+i\mu (E\rho
_{02}^{(3)}-E^{\ast }\rho _{20}^{(3)}).
\end{equation}
The spatial dc-component of $\rho _{22}^{\left( 3\right) }$ finally reads

\begin{eqnarray}
\rho _{22,dc}^{\left( 3\right) } &=&\frac{32\mu ^{2}\left( \mu ^{2}-1\right) 
}{\Delta ^{3}}{\Huge [}\left( \frac{\delta -2kv}{\left| D_{+}\right| ^{4}}%
\right) \left( \phi _{1}^{2}+\phi _{2}^{2}\right) \phi _{1}^{4}+
\label{aux3} \\
&&+\left( \frac{\delta -kv}{\left| D_{+}\right| ^{2}}\phi _{1}^{2}+\frac{%
\delta +kv}{\left| D_{-}\right| ^{2}}\phi _{2}^{2}+\frac{\delta }{\left|
D_{0}\right| ^{2}}\left( \phi _{1}^{2}+\phi _{2}^{2}\right) \right) \frac{%
\phi _{1}^{2}\phi _{2}^{2}}{\left| D_{0}\right| ^{2}}  \nonumber \\
&&+\left( \frac{\delta +2kv}{\left| D_{-}\right| ^{4}}\right) \left( \phi
_{1}^{2}+\phi _{2}^{2}\right) \phi _{2}^{4}{\Huge ].}  \nonumber
\end{eqnarray}

\section{Appendix B}

At order $\varepsilon ^{2}$ the spatially--averaged population of the
excited level, Eq.(\ref{aux2}), is 
\begin{equation}
\left\langle \rho _{22}^{\left( 2\right) }\left( v\right) \right\rangle _{z}=%
\frac{4\mu ^{2}}{\gamma \Delta _{1}^{2}}\left[ \frac{\phi _{1}^{4}}{D_{+}}+%
\frac{\phi _{2}^{4}}{D_{-}}+\frac{4\phi _{1}^{2}\phi _{2}^{2}}{D_{0}}\right]
+c.c.
\end{equation}
Now the averaging over velocities has to be carried out. As $v$ only appears
in $\rho _{22}^{\left( 2\right) }\left( v\right) $ through $D_{\pm }\left(
v\right) $ the only integrals to be done are of the type 
\begin{equation}
int_{1}=\frac{1}{\pi }\int_{-\infty }^{+\infty }d\left( 2kv\right) \frac{%
\gamma _{v}}{\gamma _{v}^{2}+\left( 2kv\right) ^{2}}\frac{\gamma }{\gamma
^{2}+\left( \delta \pm 2kv\right) ^{2}},
\end{equation}
whose result is 
\begin{equation}
int=\frac{\gamma +\gamma _{v}}{\left( \gamma +\gamma _{v}\right) ^{2}+\delta
^{2}},
\end{equation}
and thus the averaged upper level population results to be 
\begin{equation}
\left\langle \rho _{22}^{\left( 2\right) }\right\rangle =\frac{8\mu ^{2}}{%
\gamma \Delta _{1}^{2}}\left[ \frac{\gamma +\gamma _{v}}{\left( \gamma
+\gamma _{v}\right) ^{2}+\delta ^{2}}\left( \phi _{1}^{4}+\phi
_{2}^{4}\right) +4\phi _{1}^{2}\phi _{2}^{2}\frac{\gamma }{\gamma
^{2}+\delta ^{2}}\right] .  \label{rho222}
\end{equation}

At order $\varepsilon ^{3}$ the situation is similar. Now the integrals that
appear when making the velocity averaging of Eq.(\ref{aux3}) are of the type 
$int_{1}$ and also of the type 
\begin{equation}
int_{2}\left( n\right) =\frac{1}{\pi }\int_{-\infty }^{+\infty }d\left(
2kv\right) \frac{\gamma _{v}}{\gamma _{v}^{2}+\left( 2kv\right) ^{2}}\frac{%
\left( 2kv\right) }{\left[ \gamma ^{2}+\left( \delta \pm 2kv\right) ^{2}%
\right] ^{n}},  \nonumber
\end{equation}
$\left( n=1,2\right) $ whose result is 
\begin{eqnarray*}
int_{2}\left( 1\right)  &=&\frac{\gamma _{v}\delta }{\left( \gamma +\gamma
_{v}\right) ^{2}+\delta ^{2}}, \\
int_{2}\left( 2\right)  &=&\frac{\gamma _{v}\delta }{\left( \gamma +\gamma
_{v}\right) ^{2}+\delta ^{2}}\frac{\left( \gamma +\gamma _{v}\right) \left(
3\gamma +\gamma _{v}\right) +\delta ^{2}}{2\gamma ^{2}}.
\end{eqnarray*}
The final result reads

\begin{eqnarray}
\rho _{22}^{\left( 3\right) } &=&16\mu ^{2}\left( 1+A^{2}\right) \left( \mu
^{2}-1\right) \delta \left( \frac{\phi ^{2}}{\gamma \Delta _{1}}\right)
^{3}\times  \label{rho223} \\
&&\left[ \gamma ^{2}A^{2}\left( \frac{1}{\gamma _{v}\left| D_{0}\right| ^{2}}%
-\frac{1}{\gamma _{v}\left[ \left( \gamma +\gamma _{v}\right) ^{2}+\delta
^{2}\right] }+\frac{2\gamma ^{2}}{\left| D_{0}\right| ^{4}}\right) +\frac{%
2\gamma ^{2}\left( 1+A^{4}\right) \left( \gamma +\gamma _{v}\right) }{\left[
\left( \gamma +\gamma _{v}\right) ^{2}+\delta ^{2}\right] ^{2}}\right] 
{\Huge .}  \nonumber
\end{eqnarray}

{\LARGE Figure Captions}

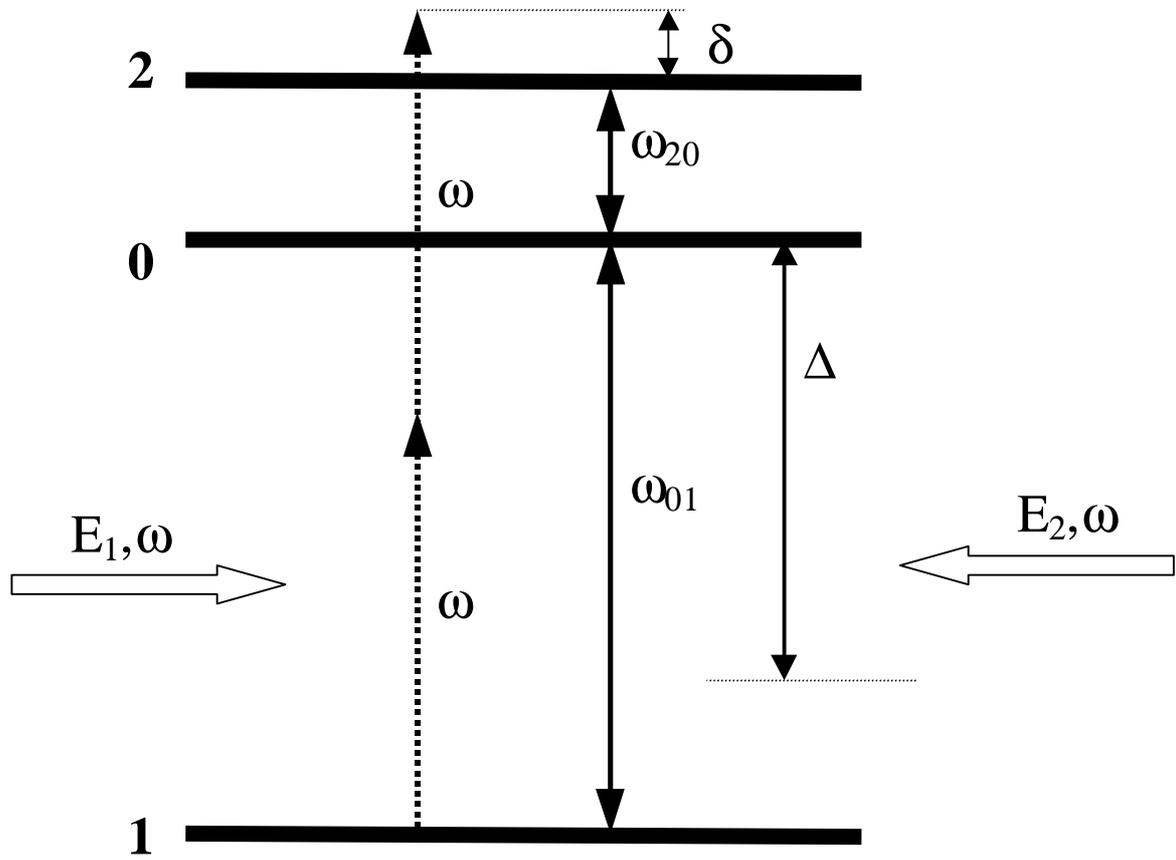
Fig.1. Energy level diagram of the three--level atoms considered in the
model. See text.

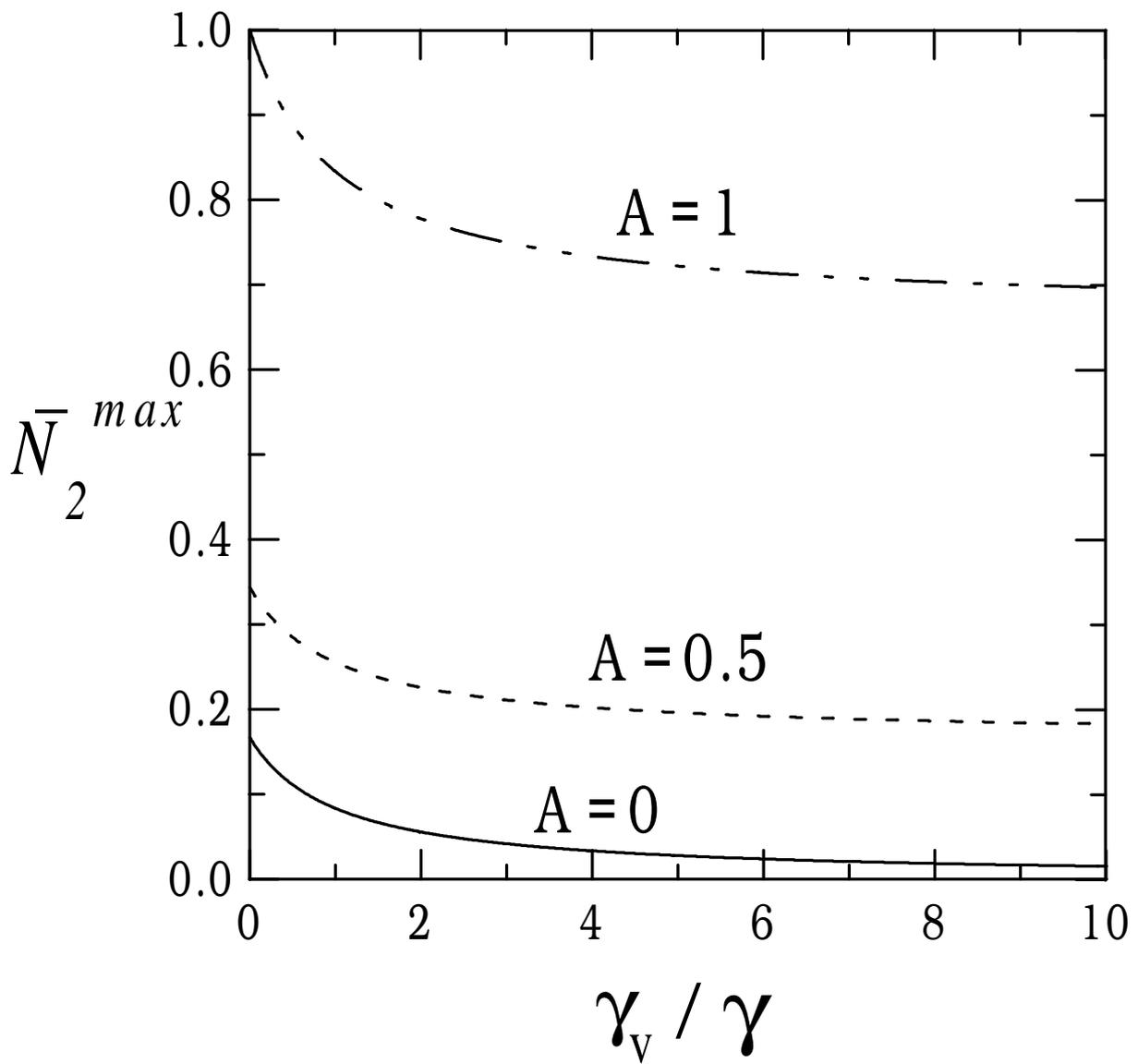
Fig.2. Maximum value of the population excited to the upper atomic level as
a function of the inhomogeneous to homogeneous width ratio $\gamma
_{v}/\gamma $ for several values of $A$. $\bar{N}_{2}^{\max }$ is $%
N_{2}^{\max }$ normalized to its maximum value (that corresponds to a
homogeneously broadened medium ($\gamma _{v}=0$) pumped by a standing wave ($%
A=1$)).

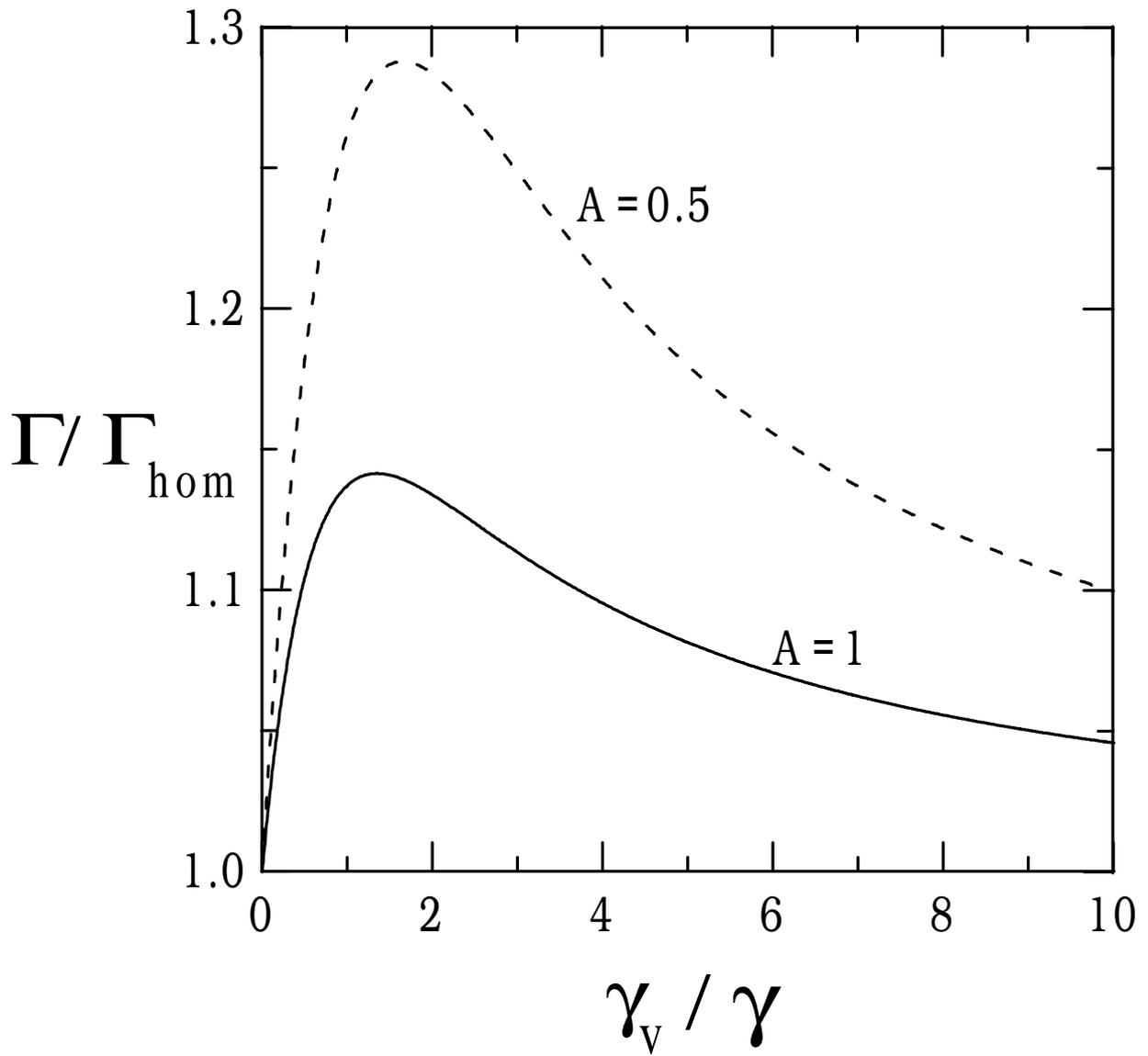
Fig.3. Width of the two photon resonance normalized to the homogeneous width
as a function of $\gamma _{v}/\gamma $ for two values of $A$. (Notice that
for a traveling wave, $A=0$, the width grows linearly with the inhomogeneous
width as $2\gamma _{v}/\gamma $.

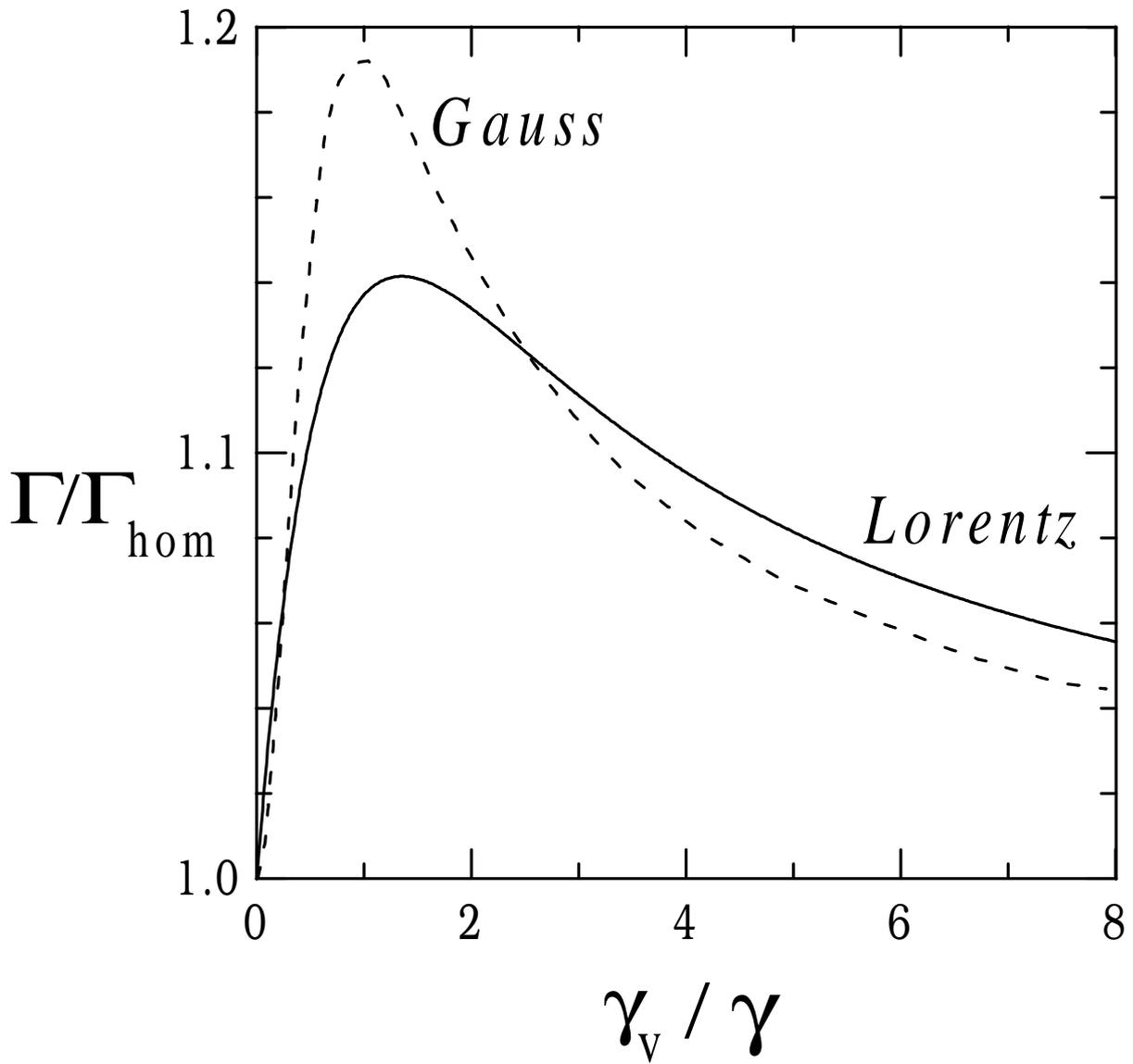
Fig.4. As in Fig.3 for a standing wave ($A=1$) for both Lorentzian (full
line) and Gaussian (dashed line) velocity distributions.

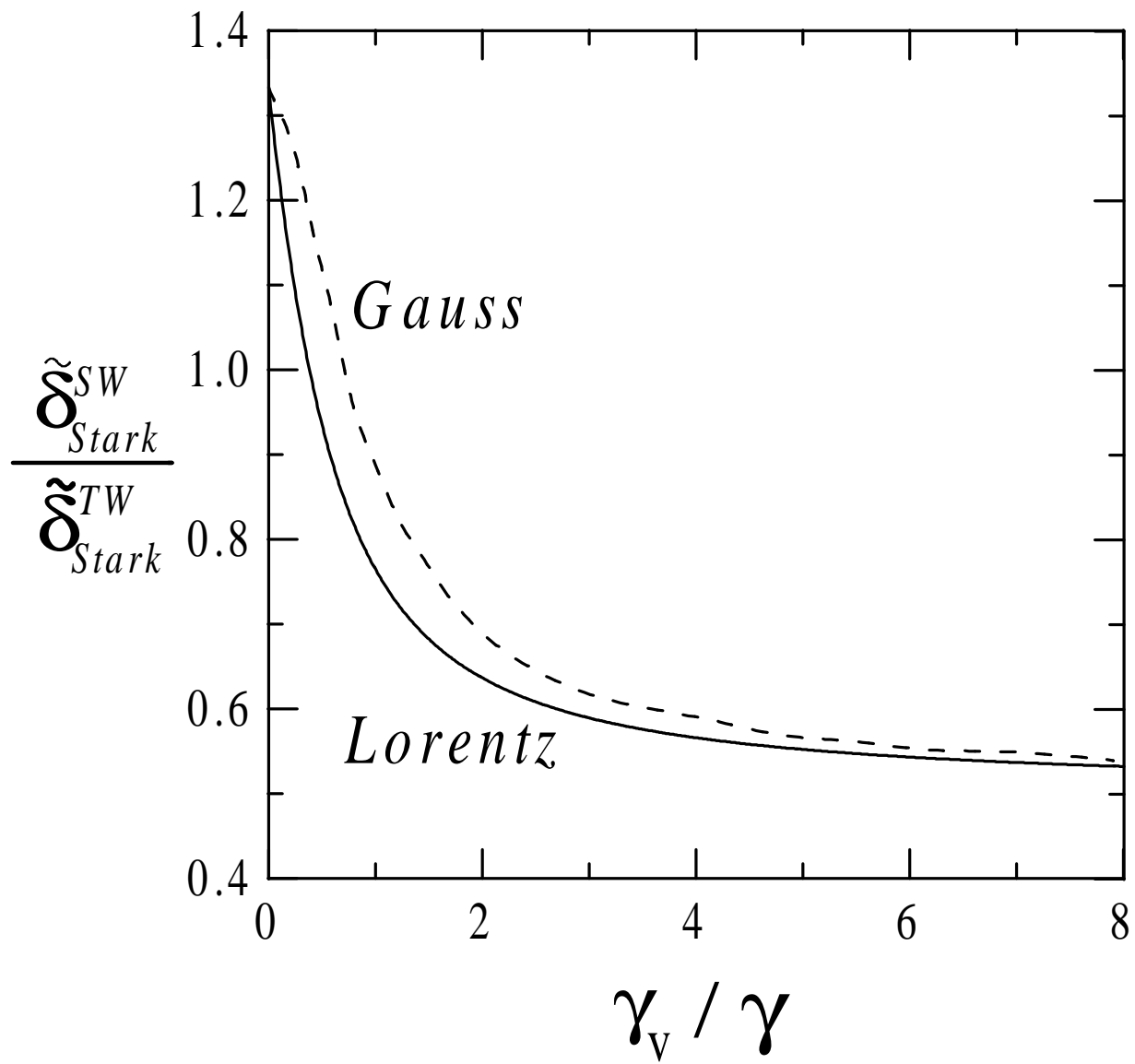
Fig.5. Dependence of the Stark shift on $\gamma _{v}/\gamma $ for a standing
wave ($A=1$) for both Lorentzian (full line) and Gaussian (dashed line)
velocity distributions.

\end{document}